\documentclass[fleqn,twoside]{article}
 \usepackage{espcrc2}

\usepackage{graphicx}
\usepackage[figuresright]{rotating}

\def\hbaromega{$h\!\!\!$\raisebox{.8ex}{-}\,$\omega$}

\def\be{\begin{equation}}
\def\ee{\end{equation}}
\def\bea{\begin{eqnarray}}
\def\eea{\end{eqnarray}}

\def\CPT{{\small $\chi$PT}}

\newcommand{\AmS}{{\protect\the\textfont2
  A\kern-.1667em\lower.5ex\hbox{M}\kern-.125emS}}

\title{Neutrino-nucleus reactions in the energy range 1-100 MeV}

\author{K. Kubodera\address{Department of Physics and Astronomy, 
       University of South Carolina \\ 
       Columbia, South Carolina 29208, USA}\thanks{Supported 
       in part by the US National Science Foundation,
                Grant Nos. PHY-0140214 and PHY-0457014;
         E-mail address: kubodera@sc.edu},
       S. Nakamura\address{Theory Group, TRIUMF\\
          4004 Wesbrook Mall, Vancouver, 
          BC V6T 2A3, Canada}\thanks{E-mail address: 
          snakamura@triumf.ca} 
          and 
       T. Sato\address{Department of Physics, Osaka University \\
        Toyonaka, Osaka 560-0043, Japan}\thanks{Supported in part
        by the Japan Society for the Promotion of Science, 
        Grant-in-Aid for Scientific Research (C) 15540275;
        E-mail address: tsato@phys.sci.osaka-u.ac.jp} }

\begin{document}

\begin{abstract}
We review some salient aspects of calculations of the neutrino-nucleus 
reaction cross sections in the low energy range (1-100 MeV).
\vspace{1pc}
\end{abstract}

\maketitle

\section{Introduction}

Neutrino-nucleus reactions in the low-energy region  
are important for multiple reasons
(see {\it e.g.,}~\cite{fy03,KN94}).
First, neutrino-nucleus reactions play important roles 
in many astrophysical processes including 
stellar nucleo-synthesis. 
Second, the observation of neutrinos emitted
in various astrophysical processes
provides valuable information in both 
astrophysical and particle-physics contexts,
and most of terrestrial experiments to measure
the astrophysical neutrinos use nuclear targets.
Third, the investigation of the neutrino properties 
with the use of accelerator neutrinos 
also very often employs nuclear targets.
All these studies call for sufficiently accurate estimates
of the cross sections for various neutrino-nucleus reactions.
It is obviously beyond the scope of this short note 
to discuss all of these individual cases.
We instead concentrate on several selected examples
which help convey the basic features of calculations
involved in low-energy neutrino-nucleus reactions.
Since the description of a
$\nu$-nucleus reaction becomes
increasingly complicated as the target mass number $A$
gets larger,
it is illuminating to first discuss
the $\nu$-$d$ reactions ($A=2$)
as the theoretically most tractable case.
We then proceed to give a brief discussion
of the $\nu$-$^{12}$C and $\nu$-$^{16}$O reactions 
and finally we add a few comments on the neutrino reactions
on medium-heavy and heavy nuclei.
We start with the explanation of two theoretical frameworks
(SNPA and EFT)
that are of general relevance in describing
electroweak processes in light nuclei.

\section{Standard nuclear physics approach (SNPA)}

In nuclear physics, the phenomenological potential 
picture has been highly successful.
In this picture an $A$-nucleon system
is described by a Hamiltonian of the form
\be
H\,=\,\sum_i^A t_i + \sum_{i<j}^A V_{ij}^{phen}+
\cdots\,, \label{HSNPA}
\ee 
where $t_i$ is the kinetic energy of the $i$-th nucleon,
$V_{ij}^{phen}$ is a phenomenological two-body potential
between the $i$-th and $j$-th nucleons,
and the dots represent potentials involving three or more 
nucleons (which are much less important than $V_{ij}^{phen}$).
The nuclear wave function $|\Psi\!>$
is obtained by solving the Shr\"{o}dinger equation
\be
H|\Psi\!>\,=\,E|\Psi\!>\,.\label{Sch}
\ee
Traditionally, finding useful truncation schemes 
for $|\Psi\!>$ has been an important branch of 
nuclear physics (the shell model, cluster model, etc.)
but, thanks to the progress in numerical computation
techniques~\cite{cs98}, it is now possible to directly solve
eq.(\ref{Sch}) for light nuclei (A$\le$11).
The phenomenological nature of $V_{ij}^{phen}$
is reflected in the fact
that its short-distance behavior is model-dependent.
(The large-distance behavior of $V_{ij}^{phen}$
is controlled by the requirement that
$V_{ij}$ should approach the Yukawa potential.)
One therefore assumes a certain functional form 
for $V_{ij}^{phen}$ and adjust the parameters appearing therein
so that the solutions of eq.(\ref{Sch}) 
for the A=2 case reproduce the two-nucleon observables.
There are by now a number of so-called 
{\it modern high-precision}
{\it phenomenological} N-N potential
that can reproduce all the existing two-nucleon data
with normalized $\chi^2$ values 
close to 1~\cite{potentials}.
To describe nuclear responses to external electroweak probes,
we use transition operators that consist of
the dominant one-body terms 
(or the impulse approximation (IA) terms)
and exchange-current (EXC) terms,
which represent the contributions
of nuclear responses involving two or more nucleons.
These transition operators 
are derived in such a manner that they are
consistent with the nuclear Hamiltonian in eq.(\ref{HSNPA})
and satisfy the low-energy theorems
and current algebra~\cite{crit}.
The theoretical framework summarized here
is a cornerstone
of contemporary nuclear physics (for a review,
see {\it e.g.,}~\cite{cs98}), and it is becoming common 
to refer to it as the {\it standard nuclear physics approach} 
(SNPA), the term apparently first used in \cite{kub01}.
SNPA has been scoring 
great phenomenological successes
in correlating and explaining a vast variety 
of nuclear phenomena~\cite{cs98}.

\section{Effective field theory (EFT)}

A new approach based on effective field theory (EFT)
is rapidly gaining ground in nuclear physics. 
The basic idea is that, in describing phenomena
characterized by a typical energy-momentum scale $Q$,
we need not include in our Lagrangian 
those degrees of freedom 
that pertain to energy-momentum scales 
much higher than $Q$;
they can be ``integrated out"
with only active low-energy degrees of
freedom (or effective fields) retained.
The effective Lagrangian, ${\cal L}_{{\rm eff}}$,
governing low-energy dynamics turns out to be given
by the sum of all possible monomials of 
the effective fields and their derivatives
that are consistent with the symmetry requirements
of the original Lagrangian.
Since a term involving $n$ derivatives
scales like $(Q/\Lambda)^n$
($\Lambda$ is a certain cutoff scale),
the terms in ${\cal L}_{{\rm eff}}$ can be organized
into a perturbative series 
in which $Q/\Lambda$ serves as an expansion parameter.
The coefficients of terms in 
this expansion scheme are called
the low-energy constants (LECs).
If all the LEC's up to a specified order $n$
can be fixed either from theory or from fitting 
to the experimental values of the relevant observables,
${\cal L}_{{\rm eff}}$ serves as 
a complete (and hence model-independent) Lagrangian
to the given order of expansion.
These considerations, applied to hadronic systems,
lead to an EFT of QCD
known as chiral perturbation theory ($\chi$PT).
A variant of $\chi$PT, called 
heavy-baryon chiral perturbation theory
(HB$\chi$PT)~\cite{JM91},
is used for a system involving a nucleon.
However, HB$\chi$PT cannot be applied 
in a straightforward manner 
to nuclei because the existence of very low-lying 
excited states in nuclei invalidates
perturbative treatments.
Following Weinberg~\cite{wei90},
we avoid this difficulty as follows.
We classify Feynman diagrams into two groups.
Diagrams in which every intermediate state
has at least one meson in flight are categorized
as irreducible, and all other diagrams 
are called reducible.
We apply the chiral counting rules
only to irreducible diagrams.
The contribution of all the two-body irreducible diagrams 
(up to a specified chiral order)
is treated as an effective potential
(to be denoted by $V_{ij}^{\mbox{\tiny{EFT}}}$)
acting on nuclear wave functions.
Meanwhile, the contributions of reducible diagrams
can be incorporated by solving the Schr\"odinger equation
\be
H^{\mbox{\tiny{EFT}}}|\Psi^{\mbox{\tiny{EFT}}}\!>\,=\,
E|\Psi^{\mbox{\tiny{EFT}}}\!>\,,
\label{Sch-EFT}
\ee
where
\be
H^{\mbox{\tiny{EFT}}}\,=\,
\sum_i^A t_i + \sum_{i<j}^A V_{ij}^{\mbox{\tiny{EFT}}}\,, 
\label{HEFT}
\ee 
We refer to this two-step procedure as
{\it nuclear} \CPT, or, to be more specific,  
{\it nuclear} \CPT\  in the Weinberg scheme.\footnote{
For an alternative form of nuclear EFT
based the KSW scheme~\cite{ksw}, see, {\it e.g.,}
Ref.~\cite{beaetal01}.}

To apply nuclear \CPT\ to a process 
that involves (an) external current(s),
we derive a nuclear transition operator 
${\cal T}^{\mbox{\tiny{EFT}}}$ 
by evaluating the complete set of 
all the irreducible diagrams
(up to a given chiral order $\nu$) 
involving the relevant external current(s).
To preserve consistency in chiral counting, 
the  nuclear matrix element of 
${\cal T}^{\mbox{\tiny{EFT}}}$ 
must be calculated with the use of nuclear 
wave functions which are governed
by nuclear interactions that represent
all the irreducible A-nucleon diagrams 
up to $\nu$-th order. 
Thus, a transition matrix in nuclear EFT
is given by
\be
{\cal M}_{fi}^{\mbox{\tiny{EFT}}}\,=\,
<\!\Psi_{\!f}^{\mbox{\tiny{EFT}}}\,|
{\cal T}^{\mbox{\tiny{EFT}}}
\,|\Psi_i^{\mbox{\tiny{EFT}}}\!>\,, \label{ME-EFT}
\ee
where the superscript ``EFT" 
means that the relevant quantities are obtained
according to EFT as described above.
If this program is carried out exactly, 
it would constitute an {\it ab initio} calculation.
In actual calculations, however, 
it is often very useful
to adopt in eq.(\ref{ME-EFT}) wave functions
obtained from SNPA. 
This eclectic approach, called hybrid EFT,
can be used for complex nuclei (A = 3, 4, ...) 
with essentially the same accuracy and ease
as for the A=2 system~\cite{pkmr,TSP-Hep}.

\section{$\nu$-$d$ reactions}

As is well known, the $\nu$-$d$ reactions
are of particular importance  
in connection with the SNO experiments~\cite{SNO}.
On the theoretical side,
because of their relative simplicity
the $\nu$-$d$ reactions 
serve as pilot cases for demonstrating the reliability
of the available calculational frameworks. 

Nakamura {\it et al.}~\cite{NSGK,Netal02} performed 
detailed SNPA calculations of the cross sections
for the reactions:
$\nu_e d\rightarrow e^-pp$,
$\nu_x d\rightarrow \nu_x pn$,
$\bar{\nu}_e d\rightarrow e^+ nn$,
$\bar{\nu}_x\rightarrow \bar{\nu}_x pn$
($x$=$e$, $\mu$ or $\tau$).
The strength of the exchange current 
(dominated by the $\Delta$-particle 
excitation diagram)
was fixed by fitting the experimental value 
of the tritium $\beta$-decay rate
$\Gamma_{\beta}^t$.
The results of Nakamura {\it et al.}
are considered to be reliable at the 1 \% level
in the solar neutrino energy range
($E_\nu\le$ 20 MeV),
and at the 5\% level up to the pion-production
threshold energy.
The basis for this statement
is as follows.
First, many electroweak observables
in light nuclei calculated with SNPA
indicate this degree of reliability of 
SNPA~\cite{cs98,Metal}.
Second, for the solar energy region,
the SNPA results are supported by EFT calculations
as well.
Butler, Chen and Kong~\cite{BCK} 
carried out an EFT calculation of the $\nu$-$d$ 
cross sections,
using the KSW scheme~\cite{ksw}.
Butler {\it et al.}'s results
(after one unknown LEC is adjusted)
are consistent with those of Nakamura et al.~\cite{Netal02}.
Further support comes from Ando {\it et al.}'s 
calculation~\cite{Aetal}.
As mentioned, hybrid $\chi$PT allows us to treat
nuclei with different mass numbers on the equal footing.
Park {\it et al.}~\cite{Petal-pp,Petal03},
applying hybrid $\chi$PT to
the GT transitions in the A=2,
3 and 4 systems, noticed that
only one unknown LEC (denoted by $\hat{d}_R$) appears,
and that $\hat{d}_R$ can be determined with good 
precision from $\Gamma_{\beta}^t$.
Hybrid $\chi$PT used in this manner
is called EFT*~\cite{MuD-ando02,kub03}.\footnote{
It is also called {\it ``more effective"} 
effective field theory (MEEFT)~\cite{TSP-Hep}.}
Ando {\it et al.}~\cite{Aetal}
used the same EFT* approach 
to carry out a parameter-free
calculation of the $\nu$-$d$ cross sections,
and their results agree with those of 
Nakamura et al.~\cite{Netal02} within 1 \%.
Thus, at least in the solar neutrino energy region, 
the SNPA calculation has
solid support from a more fundamental approach
(EFT or EFT*),
and the nature of SNPA is such that
its validity is expected to extend up 
to the pion production 
threshold  ($E_\nu\le$ 130 MeV).\footnote{
The tabulation of 
the total and differential cross sections 
for the $\nu$-$d$ reactions calculated  by
Nakamura {\it et al.}~\cite{NSGK,Netal02}
is available at:\\
$<$http://boson.physics.sc.edu/$^{\sim}$gudkov/NU-D-NSGK$>$,\\
and 
$<$http://www-nuclth.phys.sci.osaka-u.ac.jp/top$>$.}

\section{$\nu$-$^{12}$C reactions and 
$\nu$-$^{16}$O reactions}

As the mass number $A$ of the nuclear target
gets larger,
it becomes difficult to carry out full
SNPA calculations of $\nu$-$A$ reactions,
which forces us to introduce 
certain approximate treatments of the nuclear wave functions.
To illustrate issues involved in these approximations,
we discuss here the $\nu$-$^{12}$C reactions
$\nu$-$^{16}$O reactions. 
It is to be noted that $^{12}$C is an ingredient of 
liquid scintillators and $^{16}$O is the basic
component of water \v{C}erenkov counters.

The best studied method for truncating nuclear wave
functions is the shell model,
in which we introduce an average single-particle
potential $V_{ave}$ to generate single-particle orbitals
from which A-body wave functions are formed.
If we use as $V_{ave}$ the
harmonic oscillator (HO) potential (known to be
a good approximation for describing bound states
in light nuclei),
the degree of truncation of nuclear wave functions
can be specified by stating up to what excited 
HO configurations 
($n$\hbaromega \, excitations, where $\omega$
is the HO angular frequency, and $n=0,1,2,\, \ldots$) 
are included in the calculation.
With the use of a truncated model space,
the nucleon-nucleon interactions as well as
transition operators for various nuclear observables
get renormalized into effective operators,
which may be schematically denoted by
$V_{NN}^{eff}$ and ${\cal{T}}^{eff}$,
respectively.
(This is similar to what happens with the 
introduction of an EFT.)
There exist two ways to get information on 
$V_{NN}^{eff}$ and ${\cal{T}}_{eff}$.
In one method they are derived
from the ``bare" forms (pertaining to 
untruncated cases) by incorporating 
the contributions of virtual states
which lie outside the chosen model 
space~\cite{Veff}.
In the second method,
we simply list up all formally allowed terms 
for $V_{NN}^{eff}$ or ${\cal{T}}_{eff}$
and determine their prefactors
by fitting to the available experimental data.
Insofar as the number of parameters to be fixed is 
sufficiently smaller than the number of data points,
this empirical effective operator method (EEOM)
has predictive power~\cite{CK65,Wil73,BW83}.

For closed-shell light nuclei such as $^{16}$O,
one can handle a relatively large model space;
for instance, a shell-model calculation including
up to 4\hbaromega \, configurations
was reported~\cite{HJ90}.
For nuclei away from shell closures
the situation is less favorable, and
one is forced to use certain approximations 
instead of full shell-model diagonalization.
$^{12}$C belongs to this category,
and the random phase approximation (RPA)
is commonly used for this case~\cite{SJ03}.
(See however a recent no-core shell model 
calculation for the A=12 system by
Hayes {\it et al.}~\cite{Hayetal03}.)

In assessing the reliability of using the shell model
and its approximations in calculating
$\nu$-nucleus reactions, we need to pay attention 
to an apparent anomaly known for 
the inclusive reaction 
$^{12}$C$(\nu_\mu,\mu)^{12}$N*~\cite{albetal95}.
According to an LSND experiment~\cite{albetal95},
for 123.7 MeV$<$$E_\nu$$<$280 MeV,
the measured flux-averaged inclusive cross section
was more than a factor of 2 lower than that
predicted by the Fermi-gas model~\cite{GO86}\footnote{
A recent study~\cite{benetal05}
has questioned the quantitative reliability 
of the Fermi-gas even in the quasi-elastic 
scattering regime. See also~\cite{EKLV93,Nieetal05}.}
and by an RPA calculation~\cite{KLK94}. 
Although the incident neutrino energies in this experiment
lie above the energy region mentioned in the title
of this talk, 
the range of nuclear excitations involved 
in this experiment 
are relevant to lower-energy 
($E_\nu$$<$ 100 MeV) neutrino reactions as well.
It therefore concerns us
whether there is really an anomaly in
the $^{12}$C$(\nu_\mu,\mu)^{12}$N* reaction.
When the final report 
on the LSND experiment appeared~\cite{Aueetal02},
the measured flux-averaged cross section
was still significantly lower than
the existing theoretical predictions of the time.
Most recently, however, there has been 
much progress on the theoretical side.
Within RPA, Krmpoti\'{c} {\it et al.}~\cite{Krmetal02}
report that the implementation of particle-number
conservation in quasi-particle RPA
brings the theoretical prediction into agreement
with experiment.
Meanwhile, Oset and his collaborators have developed
a formalism which incorporates long-range
nuclear correlations via RPA, and which
in addition accounts for final-state interactions
and the Coulomb corrections.
Their very recent work~\cite{Nieetal05}
reports that there is no anomaly in 
the $^{12}$C$(\nu_\mu,\mu)^{12}$N* reaction.
Furthermore, in a no-core shell model 
calculation~\cite{Hayetal03}, Hayes {\it et al.}
point out that the inclusion of a realistic 
three-body interaction can remove the discrepancy 
between theory and experiment.
All these developments indicate
that the long-standing ``anomaly" is finally gone 
and that there is no basic problem with
the currently available nuclear physics approaches.

Haxton~\cite{Hax87} was the first to point out
the importance of the $\nu$-$^{16}$O reaction
in water \v{C}erenkov counters, and
he calculated the relevant cross section,
using 2\hbaromega\, shell-model wave functions
and the transition operators whose strength
are renormalized according to EEOM.
Here, the inclusive nature of the reaction 
under consideration is taken into account 
by summing all energetically allowed
final bound states of the A=16 nuclear system.
Interestingly, Kuramoto {\it et al.}'s
calculation~\cite{kuramoto90} 
based on the relativistic Fermi gas model 
(RFGM)~\cite{SM72}
indicates that the RFGM cross sections
and Haxton's results can be smoothly extrapolated
into each other at around $E_\nu=$ 60 MeV;
see also Bugaev {\it et al.}~\cite{Bugetal79}.
The most recent examples of a
calculation for the
$\nu$-$^{16}$O reaction can be found 
in~\cite{Jacetal99,Nieetal05}.

\section{Neutrino reactions on medium-heavy and heavy nuclei}

The study of low-energy neutrino reactions on
medium-heavy and heavy nuclei is of great current importance
in various contexts.
For one thing, some nuclei of this category
are either used or planned to be used
in solar neutrino experiments
with various specific features.
Furthermore, many reactions of this type play important roles
in supernova explosions and neutrino nucleosynthesis,
see {\it e.g.}~\cite{mezz}.
It is to be noted
that, depending on the condition of a core that
occurs in stellar collapse,
the average electron-neutrino energy can range typically
from 10 to 50 MeV,
and that cross sections for $\nu_e$ capture
on iron-group nuclei through A=100 are needed 
to accurately simulate core depleptonization
and to accurately determine the post-bounce initial
condition~\cite{mezz}.

Providing sufficiently accurate estimates
of the cross sections for all the  relevant
neutrino-nucleus reactions
is one of the imminent challenges facing nuclear physics.
To push ultra-large scale shell-model
calculations (or, if possible, even SNPA calculations)
for higher and higher mass numbers
is an important direction of theoretical efforts.
However there will be quite sometime before these 
{\it frontal attacks} 
can cover all the relevant nuclides.
The best strategy one could take at present is
to use a model (such as RPA or EEOM)
that can cover a wide range of nuclides
and to improve or gauge its reliability
with the help of experimental data.
For instance, Langanke {\it et al.}~\cite{Lanetal}
considered the possibility of utilizing EEOM and
high-resolution electron scattering experiments
to calibrate neutrino-nucleus cross sections
relevant to supernova neutrinos.
It is also noteworthy that there exists
an experimental project~\cite{EFR05} which, 
taking advantage of ultra-high-intensity
neutrino beams that will become available
at the Spallation Neutron Source (SNS)
at the Oak Ridge National Laboratory,
aims at measuring the cross sections of several
neutrino-nucleus reactions 
with hitherto unachievable precision.
These experiments are expected to shed much light
on the reliability and the predictive power
of nuclear models to be used for studying supernova
explosion and nucleosynthesis.

\end{document}